\documentclass{article}

\usepackage{amsmath,amssymb,pifont}
\usepackage{multicol}
\usepackage{amstext}
\usepackage{amsthm}
\usepackage{multirow}
\usepackage{booktabs}
\usepackage[skip=0pt]{subcaption}
\usepackage{times}
\usepackage{lipsum}
\usepackage[shortlabels]{enumitem}
\usepackage{cancel}
\usepackage{wrapfig}
\usepackage{array}
\usepackage{siunitx}
\usepackage{csvsimple}
\usepackage[multidot]{grffile}
\usepackage{bbm}
\usepackage{dblfloatfix}
\usepackage{hyperref}
\usepackage{makecell}
\usepackage{bbm, dsfont}
\usepackage{mathtools}
\usepackage[dvipsnames]{xcolor}
\usepackage{comment}
\usepackage[capitalise]{cleveref}

\usepackage{geometry}
\usepackage{mathpazo}
\usepackage{enumitem}

\usepackage[multiple]{footmisc}
\usepackage{mathrsfs}
\usepackage{todonotes}
\usepackage{tikz}
\usepackage{hyperref}
\usepackage{cleveref}
\usepackage{algpseudocode,algorithm,algorithmicx}
\usepackage{xfrac}

\usepackage{graphicx} 
\usepackage{color}

\usepackage{array}
\usepackage{amssymb}
\usepackage{amsmath}
\usepackage{xspace}
\usepackage{fancyhdr}
\usepackage{comment}
\usepackage{nicefrac}
\usepackage{authblk} 

\usepackage{fullpage}
\usepackage[numbers, sort, comma, square]{natbib}

\begin{document}

\title{Releasing Large-Scale Human Mobility Histograms \\ with Differential Privacy}
\author{Christopher Bian, Albert Cheu, Yannis Guzman, \\ Marco Gruteser, Peter Kairouz, Ryan McKenna, Edo Roth}
\affil{Google Research}

\maketitle

\begin{abstract}
Environmental Insights Explorer (EIE) is a Google product that reports aggregate statistics about human mobility, including various methods of transit used by people across roughly 50,000 regions globally. These statistics are used to estimate carbon emissions and provided to policymakers to inform their decisions on transportation policy and infrastructure.  Due to the inherent sensitivity of this type of user data, it is crucial that the statistics derived and released from it are computed with appropriate privacy protections.  In this work, we use a combination of federated analytics and differential privacy to release these required statistics, while operating under strict error constraints to ensure utility for downstream stakeholders.  In this work, we propose a new mechanism that achieves $ \epsilon \approx 2 $-DP while satisfying these strict utility constraints, greatly improving over natural baselines.  We believe this mechanism may be of more general interest for the broad class of group-by-sum workloads.  
\end{abstract}

\section{Introduction}

Publishing statistics derived from human mobility can help tackle the world's sustainability challenges. Google's Environmental Insight Explorer (EIE)\cite{eieblog} provides cities with emissions data to help them set emissions reduction targets and inform decarbonization strategies~\cite{cop26blog}. In its transportation section, EIE shares estimates of transportation greenhouse gas emissions within a geographic region of interest (city, county, etc.) derived from aggregated Google Location History smartphone users who have enabled the location history setting (which is off by default). An on-device\cite{odlhblog} Tensorflow model takes features related to the location data for a given trip and predicts a classification label that represents the mode of transit. These include passenger vehicle, motorcycle, walking, cycling, as well as public transit such as bus, tram, rail, subway, and ferry.

To view data on EIE, visit \url{http://insights.sustainability.google} where you can search for or browse the cities who have made their data publicly available on the coverage map. Over 2,400 cities (regions) have published their EIE data to the public, which consist of transportation mode share, and their associated trip count, distance traveled, and estimated CO2e emissions. These are the primary statistics which have been made differentially private. For non-public regions, only a small sample of the differentially private transport data are available, consisting only of the distribution of mode share per year.

To gain access to the data, you must be associated with the city in an official capacity, by requesting an Insights Workspace account. Upon successful validation of city affiliation, differentially private transportation statistics, as well as other available datasets such as solar potential, building emissions, etc., are released for your approved regions (click the link "sign up to access" on the upper right hand corner).

Cities use this data as a benchmark to reference in annual greenhouse gas inventories and to inform goals and strategies in climate action plans. In order to do this for mobility related initiatives, city-users will typically look at mode-share, which refers to the distribution of the methods of travel across the region. The tool also provides users a way to assess the CO$_2$ impact of various decarbonization strategies by modifying the distances traveled (commonly referred to as VKT or VMT) and their associated fuel efficiency and carbon intensity coefficients to represent different scenarios. For example, a city-user may test a vehicle-to-public-transit policy by moving 1 million kilometers from private passenger vehicles into transit buses. When accounting for the high-occupancy characteristics for their city's transit system, they may find this only requires 50,000 additional bus miles which can be adjusted in EIE under that mode. Cities may continue to layer on more scenarios, such as upgrading 50\% of the bus fleet to electric buses, which would result in halving the carbon intensity coefficient. Another scenario could be to further upgrade 25\% of the bus fleet to more fuel efficient drive-trains by modifying the fuel efficiency coefficient. EIE will recompute emissions for the entire city as these adjustments are made, and present a new emissions estimation which can be saved for sharing with stakeholders during their policy planning journey.

Given the private nature of the underlying data, we explore the feasibility of deriving and sharing such data even more privately with differential privacy (DP) while still maintaining utility for the stakeholders who will receive these statistics and use them to make city planning decisions like setting targets for cycling, electrification, or facilitating mode shifts to public transit.

We further seek to enable a federated analytics (FA) approach \cite{ramage2021federated} to processing the data, such that single user data points are not accessible by the analytics service provider, only aggregates. We do not focus on the FA architecture in this work, but instead focus on the challenge of implementing a DP mechanism on top of it. The differential privacy techniques deployed into the FA infrastructure are actively being used to collect data in production for EIE, with the intention for the annual release of 2024 mobility and emissions data in the late spring of 2025.

DP histograms over location data have been published before \cite{aktay2020google,Wellenius_2021,adeleye2023publishing}, but our setting has a unique combination of features. First, the dimensionality is massive: over 4 \textit{million} total statistics need to be released. Second, we assess utility with an application and domain specific relative weighted error function, different than what is typically studied in the literature. Third, there is a large variance in the magnitude of user contributions. Unlike other large-scale DP efforts, like the 2020 census \cite{abowd20222020}, which largely computes counts over single rows of data from users, in this work we compute sums over metrics like ``total distance,'' which can vary dramatically depending on modes of transportation (consider the difference between a user taking a walk in their local park versus taking a cross-continent flight).  Because of this structural variance, and because users can contribute multiple rows of information, it is difficult to bound user contribution and scale the resulting sensitivity (and thus the noise applied), in a way that maintains utility of large-magnitude metrics without drowning out those of small-magnitude metrics. To our knowledge, there are no published techniques we could use for this scenario. 

Our resulting approach relies on the use of local activity-level scaling norms, which scale data on user devices before their collection, depending on the activities (modes of transportation) taken. We then utilize a single clipping bound across a user's entire contribution, computed to take into account their re-scaled contributions. These clipped contributions are aggregated securely under federated analytics to a central server, where we perform server-side rescaling (to readjust the data), noise addition (to provide DP guarantees), and an additional post-processing thresholding step to discard over-noisy data. We evaluate our algorithm on a dataset with hundreds of millions of users, and find that we can achieve a DP guarantee of $\epsilon \approx 2 $ for a privacy unit of (user, week), with future work potentially bringing this close to $\epsilon \approx 1$ for the same privacy unit.





\section{Problem Setup}

\paragraph{\textbf{Data}}

Each user device contains a collection of records, and each record corresponds to a trip.  Each record has three categorical attributes: region (e.g., New York City, Tashkent, Santiago), activity type (e.g, cycling, driving, subway), and trip direction (within region, outbound, and inbound).  There are $\approx 50,000$ possible regions, 9 activity types, and 3 trip directions.  Each record also has two numerical attributes: distance traveled (measured in kilometers) and duration of the trip (measured in seconds).

Server-side, each record is associated with exactly one ISO-8601 Monday-Sunday week. We will later explain how this is used with respect to the differential privacy guarantee. In addition to ensuring DP outputs, the system ensures raw server-side data and intermediate aggregates are only accessible within a time window of approximately one month. 

There are hundreds of millions of users contributing data every week, and billions of total records, each consisting of one trip from a user.  Each user can contribute a different number of records, as well as trips with different magnitudes, and the contributions are difficult to bound a priori. A comparable server-side proxy dataset was used to tune our parameters and for evaluation of utility and privacy loss. 

\paragraph{\textbf{Data Limitations}} The aggregated mobility data underlying this study has the following limitations. First, it is limited to Android users who have turned on the Google Location History service, which is off by default. The sample size and representativeness of the data therefore varies across regions. Second, location is coarsened to regions, typically to cities or larger administrative regions. As the trip counts, travel distances, and trip duration statistics are only approximately allocated to these coarse `regions', this may not be fully representative of precise location. No individual user data was manually inspected during this study, only highly aggregated statistics over large populations were handled for the purpose of evaluating differential privacy mechanisms.


\paragraph{\textbf{Workload}}

Every week $w$, we aim to release statistics about user data associated with that week (denoted $D_w$). Specifically, for each (region, direction, activity) combination, we would like to release three \emph{metrics} of interest: the number of trips, aggregate trip distance, and aggregate trip duration. This amounts to releasing three histograms (one per metric), each containing $50K \times 3 \times 9 = 1.35$ million partitions.  These statistics are post-processed by applying vehicle fleet distributions and their fuel carbon intensities in order to estimate CO$_2$ equivalent emissions per region, then surfaced to users on EIE for further analysis and use.

\paragraph{\textbf{Evaluation Criteria}}

Given noisy answers to the workload queries, we judge their utility according to their ``weighted relative error.''  For a given metric and each (region, direction, activity) we calculate the relative error between the true value and noisy estimate.  We then take a weighted average of this quantity across the entire histogram, ignoring entries where fewer than $2000$ client devices contributed data.  The weight for a given region $r$, direction $d$ and activity $a$ is defined as $n_{r, d, a} / n_r$, where $n$ is the total number of trips for a given setting.  This weights entries according to how important they are within the given region.  

In this work, we take a utility-first approach\footnote{See "Data Limitations" and "Privacy Goals" subsections for how we limit the privacy risk and justify this utility-first approach within the confines of strong privacy guarantees.}, as the end application requires an average relative weighted error of $\approx 3\%$  to be useful in downstream applications.

The use of relative error is common in production. For example, Adeleye et al. assess quality of released Wikipedia statistics by computing their ratios against ground truth (refer to ``Relative error distribution'' in their Section 3.4) \cite{adeleye2023publishing}. We note that they do not use weighting similar to ours.


\paragraph{\textbf{Privacy Goals}} Our privacy goals are twofold. First, the raw user trip data should stay on users' respective devices: for a given week $w$, the dataset $D_w$ is distributed across users and the server should only see aggregates across many users.  This first goal is achieved using federated analytics \cite{ramage2021federated}. Second, the aggregates should be appropriately randomized before release to downstream stakeholders to provide a suitable level of formal differential privacy (DP).

To achieve the second goal, we begin by specifying the unit of information that our algorithm will protect. This is done by defining an adjacency relation between datasets. In our setting, a user contributes trip data over time so it is meaningful to define adjacency in the following way: $D_w$ and $D'_w$ are adjacent when $D'_w$ can be formed by adding or removing one user's trips to $D_w$. Thus, the \emph{privacy unit} is (user, week) addition.\footnote{We also considered other units of privacy, including (user, activity, week) and (user, metric, week) --- it is trivial to convert guarantees over these finer-grained units to (user, week), however.}

Now that we have defined the privacy unit, we define the DP objective. For any pair of adjacent datasets $D_w,D'_w$ and any given event $E$, an $\epsilon$-DP algorithm $M$ guarantees the odds of the event changes by at most an $\exp(\epsilon)$ factor:
$$
\mathrm{Pr}[M(D_w) \in E] \leq \exp(\epsilon)\cdot \mathrm{Pr}[M(D'_w) \in E]
$$
Intuitively, $M$ will add enough noise to its computation to hide which of the two datasets was given as input; Alice is assured that the distribution of the algorithm's output will look roughly the same regardless if she went to Milan during $w$ or stayed in Milwaukee in the same time window.

\section{Mechanism Design} \label{sec:mechanism_design}

In this section, we discuss some of the challenges of this task, our key insights, and how these informed our mechanism design decisions.  

\paragraph{\textbf{Key Primitive: The Laplace Mechanism}}
We begin by describing a key primitive that will be shared by all of the mechanisms we consider in this section: the Laplace mechanism \cite{DMNS}.  Given a vector-valued contribution for each user $\mathbf{v}_1, \dots, \mathbf{v}_n$, and an $L_1$ clipping bound $C$, the Laplace mechanism releases 
$$\mathbf{\tilde{v}} = \sum_{i=1}^n \text{clip}(\mathbf{v}_i, C) + \text{Lap}\Big(0, \frac{C}{\epsilon}\Big)$$

where $\text{clip}(\mathbf{v}_i, C) = \mathbf{v}_i \text{min}(1, \nicefrac{C}{|| \mathbf{v}_i} ||_1)$.
This mechanism is known to provide $\epsilon$-DP.  $C$ is a crucial hyperparameter that must be carefully chosen to balance the bias from clipping and the variance due to Laplace noise.  We know from prior work that choosing $C$ based on an appropriate quantile (e.g., 95\%) of $|| \textbf{v}_i ||_1$ typically provides good utility \cite{liu2023algorithms,andrew2021differentially}.  This quantile can be calculated privately using e.g., the exponential mechanism.  Alternatively, $C$ can be tuned via a grid search using proxy data which will not require DP protections.

\paragraph{\textbf{Baseline Approaches}}

There are two obvious ways one can invoke the Laplace mechanism primitive to solve this problem.

\begin{enumerate}
    \item \textbf{Budget Split:} Privately compute the histogram over (region, direction) for each (activity, metric), using a privacy budget of $\dfrac{\epsilon}{27}$ for each one.
    \item \textbf{Joint Clipping:} Privately compute a single global histogram with an entry for each (region, direction, activity, metric).
\end{enumerate}

There are some interesting trade-offs between these two baseline approaches.  Recall that the evaluation metric is a weighted relative error, and therefore histogram entries with smaller values should also have smaller error.  The first approach invokes the Laplace mechanism primitive multiple times and therefore must split the privacy budget between each invocation.  The benefit of this approach is that the clipping threshold can be tuned separately for each histogram, which means the noise can be calibrated approximately to the magnitude of the entries in that histogram, which can improve relative error.  The drawback of this approach is that the privacy budget must be divided according to the number of histograms answered privately (27 in this case).  The second approach does not require dividing the privacy budget many ways, but it adds i.i.d. noise to the histogram, and hence noise is not calibrated to the magnitude of the histogram entries like the first approach.

\paragraph{\textbf{Our Approach: Activity+Metric Scaling}}

In this section, we develop a method that offers the benefits of each approach above through a careful rescaling of the histogram before clipping and adding noise.  
We make the following key observations:

\begin{enumerate}
    \item The histogram for different (activity, metric) pairs can vary by multiple orders of magnitude.  For example, the distance traveled by walking is on the order of kilometers, while the distance traveled by flying can be on the order of hundreds or thousands of kilometers.  Moreover, values for different metrics have different units, and depending on the units used, could vary significantly.  E.g., If distance is measured in kilometers and duration is measured in seconds, these entries of the histogram corresponding to duration will have the largest magnitude and hence enjoy the lowest relative error.  
    \item Users typically only contribute data to a small set of regions, and hence it does not make much sense to split the queries up by region.  Similarly, different users may contribute primarily to different activities, and hence it can be helpful to consider different activities together in the same query.  
    \item As noted above, the clipping threshold is often calibrated to the \emph{outlier} users (via the 95\% quantile), to ensure we preserve most of the signal and do not introduce too much bias from clipping.  If the outlier users for each activity are different, then we can clip the joint histogram across all activities (appropriately scaled) without increasing the clipping threshold too much.
\end{enumerate}

\begin{figure*}[t]
\includegraphics[width=\textwidth]{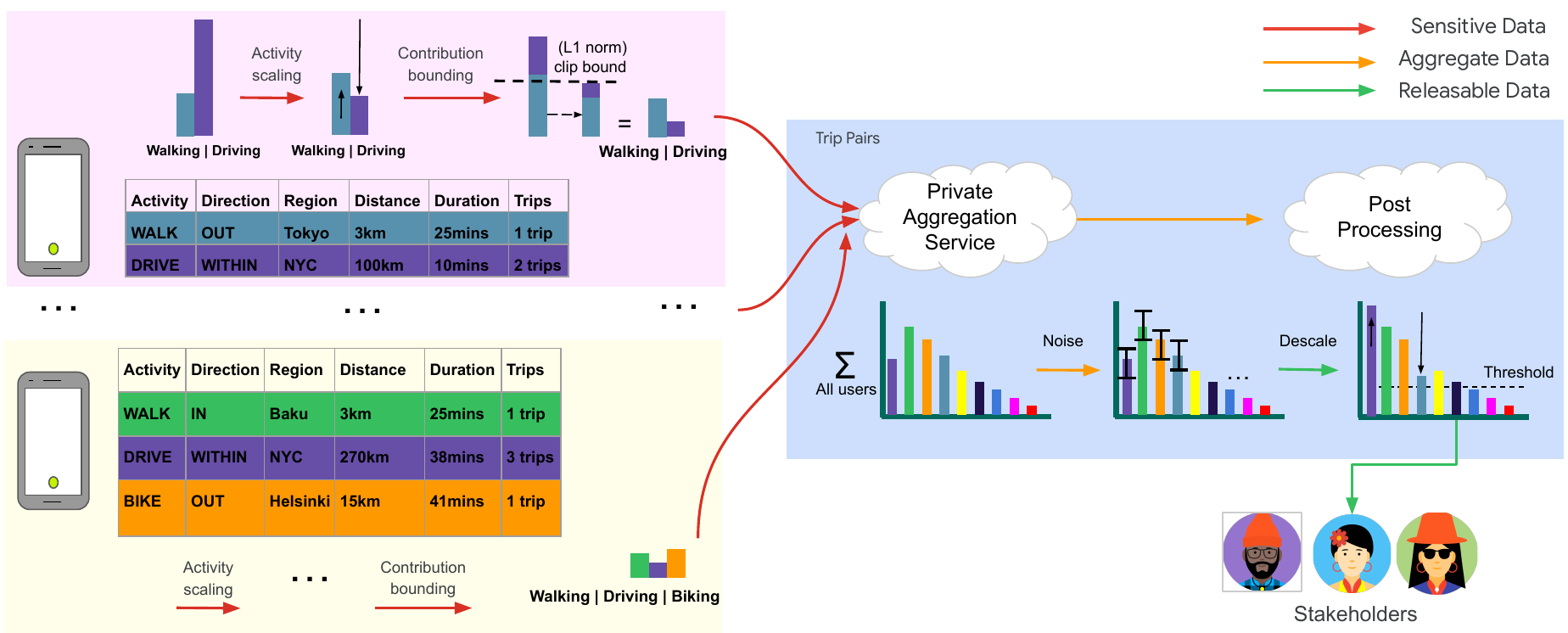}
\caption{A complete overview of the data collection and processing steps, including the ``Activity + Metric Scaling Mechanism'' (with two sample devices shown).  Data starts on device, is scaled and clipped locally, before being aggregated by the Private Aggregation Service. Once aggregated on the server, we add Laplace noise centrally, and perform post-processing steps of descaling and thresholding, before releasing the private data to downstream emissions calculations and then to relevant stakeholders.} \label{fig:mechanism}
\end{figure*}

\begin{algorithm}
\caption{ClientWork}
\label{alg:clientwork}
\begin{algorithmic}
\Require scales $S \in \mathbb{R}^{9 \times 3}$, clip $C \in \mathbb{R}$, user $i \in \mathbb{N}$ 

\State $\mathbf{v} \gets \mathbf{0} \in \mathbb{R}^{9 \times 3 \times 50K \times 3}$
\For{activity $a$, region $r$, direction $d$, distance, duration \textbf{in} records$_i$}
    \State $\mathbf{v}(a, \text{`num trips'}, r, d) \mathrel{+}= 1$ / $S(a, \text{`num trips'})$
    \State $\mathbf{v}(a, \text{`distance'}, r, d) \mathrel{+}= $ distance / $S(a, \text{`distance'})$
    \State $\mathbf{v}(a, \text{`duration'}, r, d) \mathrel{+}= $ duration / $S(a, \text{`duration'})$
\EndFor
\State \Return $\mathbf{v} \cdot \min(\frac{C}{|| \mathbf{v} ||_1}, 1)$
\end{algorithmic}
\end{algorithm}

\begin{algorithm}
\caption{ServerWork}
\label{alg:serverwork}
\begin{algorithmic}
\Require scales $S \in \mathbb{R}^{9 \times 3}$, clip $C \in \mathbb{R}$, privacy budget $\epsilon$

\For{User $i$}
    \State  $\mathbf{v}_i \leftarrow$ ClientWork($S$, $C$, $i$) \Comment \textcolor{orange}{\textbf{Client}}
\EndFor
\State $\mathbf{v} \leftarrow$ SecureSum($\mathbf{v}_i$) \Comment{\textcolor{orange}{\textbf{Client $\rightarrow$ Server}}}
\State $\tilde{\mathbf{v}} = \mathbf{v} + \text{Lap}\Big(\frac{C}{\epsilon} \Big)^{9 \times 3 \times 50K \times 3}$ \Comment \textcolor{orange}{\textbf{Server}}

\For{\textbf{each possible} activity $a$, metric $m$, region $r$, direction $d$}
    \State $\tilde{\mathbf{h}}(a, m, r, d) = \tilde{\mathbf{v}}(a, m, r, d) S(a, m)$  \Comment \textcolor{orange}{\textbf{Server}}
\EndFor
\State \Return $\tilde{\mathbf{h}}$
\end{algorithmic}
\end{algorithm}

These observations lead the ``Activity+Metric Scaling Mechanism'', summarized in \cref{fig:mechanism} and described more precisely in \cref{alg:clientwork,alg:serverwork} and below.  We will use the notation $\mathbf{v}(a, m, r, d)$ to index into a histogram defined over activities $a$,  metrics $m$, regions $r$, and directions $d$.

\begin{enumerate}
    \item Let $S(a, m)$ be the 95\% quantile of $|| \mathbf{h}_i(a, m, \cdot, \cdot) ||_1$, where $\mathbf{h}_i(a, m, \cdot, \cdot)$ is the subvector of $\mathbf{h}_i$ corresponding to activity $a$ and metric $m$.
    \item Scale each user contribution: Now let $\mathbf{v}_i(a, m, r, d) = \nicefrac{\mathbf{h}_i(a, m, r, d)}{ S(a, m) }$.
    \item Invoke the Laplace mechanism primitive on $\sum_i \mathbf{v}_i$ to get $\tilde{\mathbf{v}}$.
    \item Apply the reverse scaling and return $\tilde{\mathbf{h}}(a, m, r, d) = \tilde{\mathbf{v}}(a, m, r, d) S(a, m)$.
\end{enumerate}

By scaling each value by the inverse of the 95\% quantile for that activity and metric, we are essentially transforming the values to a new domain where they have similar magnitudes.  This allows us to naturally handle the different scales across activities and metrics without splitting the privacy budget.

This approach can be particularly effective if the outlier users for each (activity, metric) are different.  This is reasonable across activities, as there may be many users who are extreme cyclists, runners, flyers, drivers, etc., but few users who are extreme in multiple categories.  However, across metrics, this is likely to be less true as e.g., duration and distance traveled are highly correlated.  Hence, there is less to be gained by considering the the metrics jointly instead of simply splitting the privacy budget across them. 

\section{Empirical Evaluation}

In this section, we evaluate three different mechanisms for privately estimating the EIE histograms, the two baseline approaches described in \cref{sec:mechanism_design} as well as our new approach.  These experiments were run on our server-side proxy dataset, which allowed us to compare mechanisms efficiently, and estimate how well they might do with different privacy budgets before deployment in production.  

We vary the privacy budgets, and include a reference line for our target $\approx 3\%$ relative error.

\paragraph{\textbf{Analysis of Results}}


\begin{figure}
\centering
\hspace{-5em}
\begin{subfigure}{0.45\textwidth}  
    \centering
    \includegraphics[width=\textwidth]{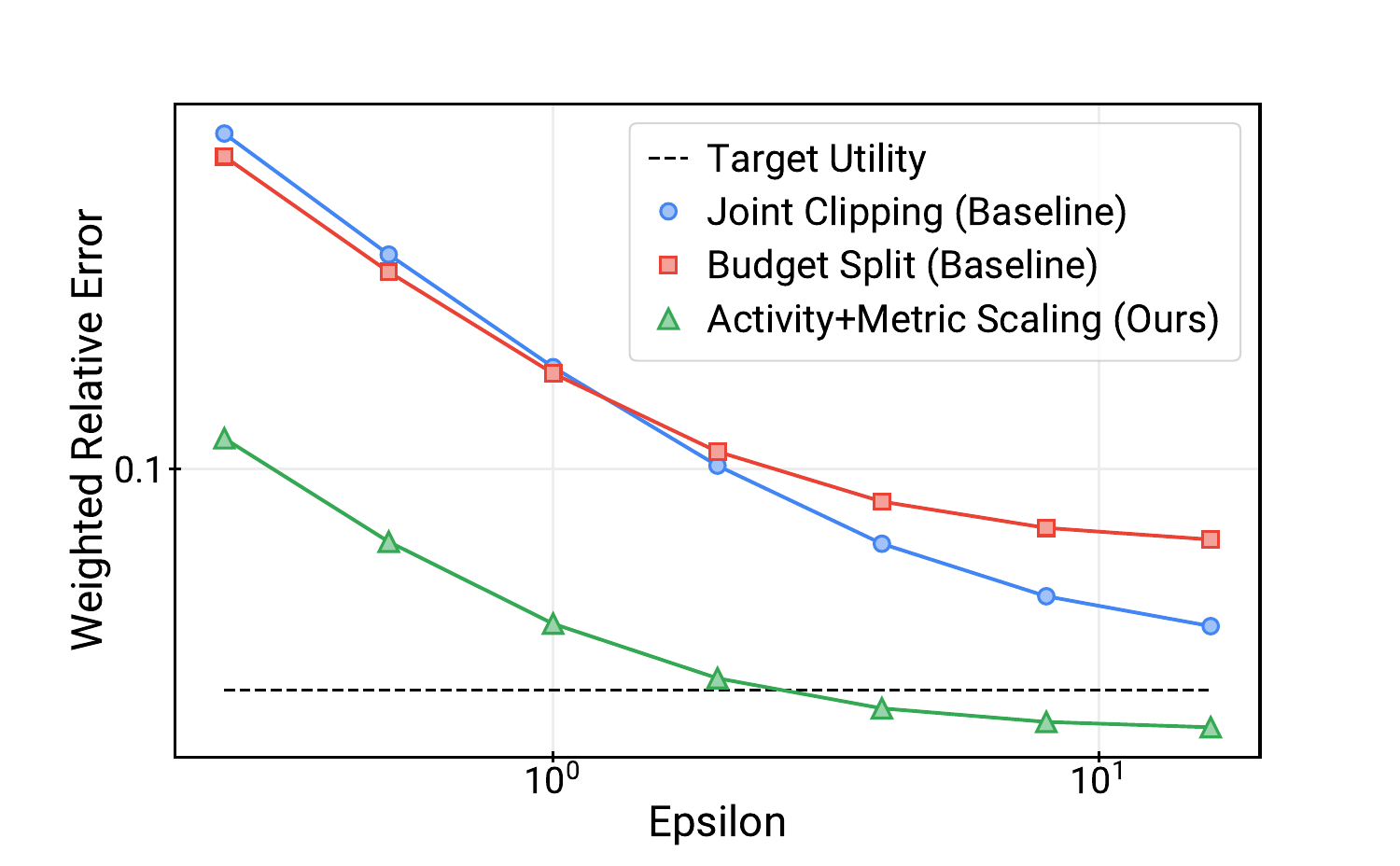}
    \caption{\label{fig:plot}}
\end{subfigure}
\begin{subfigure}{0.45\textwidth}
    \centering
    \vspace{2.5em}
    \begin{tabular}{c|ccc}  
        \textbf{Mechanism} & \multicolumn{3}{c}{\textbf{Weighted Relative Error}} \\
        ($\epsilon = 2$) & Num Trips & Distance & Duration \\\hline
        Joint Clipping & 0.195 & 0.072 & 0.038 \\
        Budget Split & 0.091 & 0.150 & 0.088 \\
        Activity + Metric & 0.028 & 0.040 & 0.028 \\
        Scaling & & & \\
    \end{tabular}
    \vspace{3em}
    \caption{\label{fig:table}}
\end{subfigure}
\caption{(a) overall weighted relative error of two baseline mechanisms and our mechanism for varying privacy budgets.  (b) Error breakdown of each mechanism for each metric at $\epsilon=2$.}
\label{fig:experiments}
\end{figure}

As shown in \cref{fig:plot}, among the two baselines discussed in \cref{sec:mechanism_design}, the Budget Split baseline was slightly better for smaller privacy budgets, while the Joint Clipping baseline was better for larger privacy budgets.  Neither baseline met the utility goal of $3\%$ weighted relative error even for $\epsilon = 16$.  Our new Activity + Metric Scaling mechanism performed much better across the board, and came close to the utility target for $\epsilon = 2$.  

As shown in \cref{fig:table}, the Joint Clipping baseline enjoyed 5$\times$ lower error on the ``duration'' metric than the the ``num trips'' metric.  This is likely due to the fact that the duration (measured in minutes) of most trips is much greater than $1$, so those entries of the histogram are much larger than the entries corresponding to ``num trips''.  The Budget split baseline has similar error between ``duration'' and ``num trips'' but a 2$\times$ worse error for the ``distance'' metric.  We believe this might be because the distance traveled is a higher variance quantity and we lose significant signal from the outlier users above the 95\% quantile.  Our new mechanism achieves similar error between all three metrics, although error on the ``distance'' metric is still somewhat larger.


\section{Lessons Learned and Future Work}


In this work, we developed a new mechanism for privately computing histograms under user-level DP, when different subvectors have different scales. On top of the protections guaranteed by federated analytics (data being highly aggregated, and locations coarsened to regions), we applied this mechanism to Environmental Insights Explorer, and  were able to achieve a $\epsilon \approx 2$ (user,week)-level-DP guarantee while meeting strict utility requirements.  This guarantee further protects user anonymity and obscures fine detail. 

We believe the mechanism we proposed in this work may be more broadly applicable in other domains.  However, some work is needed to generalize the approach and streamline the process.  One limitation of our approach is that we use a proxy server side dataset to find the appropriate quantiles and tune the clipping thresholds.  Developing an automated approach to setting the hyperparameters without public data remains an important question.  

Another interesting direction for future work is to design mechanisms that adapt intelligently to the input workload.  The mechanism we propose here does very little to exploit knowledge of the workload, other than the fact that is a weighted average of relative errors.  It would be interesting to incorporate query importance into the scaling parameters, as well as the magnitude of the values themselves.

\section{Acknowledgments}
We would like to thank Brendan McMahan for his support throughout this work and for his insightful comments on earlier drafts of this paper. 

\newpage
\bibliographystyle{acm}
\bibliography{ref}
\end{document}